\documentclass[times,twocolumn,final]{elsarticle}

\usepackage{medima}
\usepackage{framed,multirow}

\usepackage{amssymb}
\usepackage{latexsym}

\usepackage{url}
\usepackage{multirow}
\usepackage[table]{xcolor}
\usepackage{soul,color}
\usepackage{amssymb}
\usepackage{bbm}
\usepackage{bm}
\usepackage{soul}
\usepackage{booktabs}
\usepackage{mathtools}
\usepackage{arydshln}

\usepackage{amsmath}
\usepackage{pifont}
\newcommand{\cmark}{\ding{51}}
\newcommand{\xmark}{\ding{55}}

\definecolor{red}{rgb}{.9,0.1,0.1}
\definecolor{blue}{rgb}{.2,0.5,0.7}

\usepackage{hyperref}

\definecolor{newcolor}{rgb}{.8,.349,.1}

\journal{Medical Image Analysis}

\begin{document}

\verso{Stine Hansen \textit{et~al.}}

\begin{frontmatter}

\title{Anomaly Detection-Inspired Few-Shot Medical Image Segmentation Through Self-Supervision With Supervoxels\tnoteref{tnote1}}%
\tnotetext[tnote1]{All the authors are with the UiT Machine Learning Group (\url{machine-learning.uit.no}) and with  \emph{Visual Intelligence}, a Norwegian Centre for Research-based Innovation (\url{visual-intelligence.no}).}

\author[1]{Stine \snm{Hansen}\corref{cor1}}
\cortext[cor1]{Corresponding author}
\ead{s.hansen@uit.no}
\author[1]{Srishti \snm{Gautam}}
\ead{srishti.gautam@uit.no}
\author[1]{Robert \snm{Jenssen}}
\ead{robert.jenssen@uit.no}
\author[1]{Michael \snm{Kampffmeyer}}
\ead{michael.c.kampffmeyer@uit.no}

\address[1]{Department of Physics and Technology, UiT The Arctic University of Norway, NO-9037 Tromsø, Norway}

\received{22 June 2021}
\accepted{1 February 2022}
\availableonline{11 February 2022}

\begin{abstract}
Recent work has shown that label-efficient few-shot learning through self-supervision can achieve promising medical image segmentation results. However, few-shot segmentation models typically rely on prototype representations of the semantic classes, resulting in a loss of local information that can degrade performance. This is particularly problematic for the typically large and highly heterogeneous background class in medical image segmentation problems. Previous works have attempted to address this issue by learning additional prototypes for each class, but since the prototypes are based on a limited number of slices, we argue that this ad-hoc solution is insufficient to capture the background properties. Motivated by this, and the observation that the foreground class (e.g., one organ) is relatively homogeneous, we propose a novel anomaly detection-inspired approach to few-shot medical image segmentation in which we refrain from modeling the background explicitly. Instead, we rely solely on a single foreground prototype to compute anomaly scores for all query pixels. The segmentation is then performed by thresholding these anomaly scores using a learned threshold. Assisted by a novel self-supervision task that exploits the 3D structure of medical images through supervoxels, our proposed anomaly detection-inspired few-shot medical image segmentation model outperforms previous state-of-the-art approaches on two representative MRI datasets for the tasks of abdominal organ segmentation and cardiac segmentation.
\end{abstract}


\end{frontmatter}


\section{Introduction}
Many applications in medical image analysis, such as diagnosis~\citep{tsochatzidis2021integrating}, treatment planning~\citep{chen2021deep}, and quantification of tissue volumes~\citep{abdeltawab2020deep} rely heavily on semantic segmentation. To lessen the burden on the medical practitioners performing these manual, slice-by-slice segmentations, the use of deep learning for automatic segmentation has a great potential. Unfortunately, existing segmentation frameworks~\citep{ronneberger2015u,li2018h,isensee2021nnu} depend on supervised training and large amounts of densely labeled data, which are often unavailable in the medical domain. Moreover, their generalization properties to previously unseen classes are typically poor, necessitating the collection and labeling of new data to re-train for new tasks. Due to the huge number of potential segmentation tasks in medical images, this makes these models impractical to use. 

Inspired by how humans learn from only a handful of instances, few-shot learning has emerged as a learning paradigm to foster models that can easily adapt to new concepts when exposed to just a few new, labeled samples. These models typically follow an episodic framework~\citep{vinyals2016matching} where, in each episode, $k$ labeled samples, called the support set, are used to segment the unlabeled query image(s). The models are trained on one set of classes and learn to, with only a few annotated examples, segment objects from new classes. A \textit{trained} few-shot segmentation (FSS) model is thus able to segment an unseen organ class based on just a few labeled instances. However, in order to avoid over-fitting, typical FSS models rely on training data containing a large set of labeled training classes, generally not available in the medical domain.  

In a recent work, \cite{ouyang2020self} proposed a label-efficient approach to medical image segmentation, building on metric-learning based prototypical FSS~\citep{liu2020part,wang2019panet}. They suggest a model that follows the traditional few-shot episodic framework, where class-wise prototypes are extracted from the labeled support set and used to reduce the segmentation of the unlabeled query image to a pixel-wise prototype matching in the embedding space. Whereas traditional few-shot learning models require a set of annotated training classes, \cite{ouyang2020self} propose a clever way to bypass this need by employing self-supervised training~\citep{jing2020self}. Instead of sampling labeled support and query images, they construct the training episodes based on \textit{one} unlabeled image slice and its corresponding superpixel~\citep{ren2003learning} segmentation: One randomly sampled superpixel serves as foreground mask, and together with the original image slice, these form the support image-label pair. The query pair is then constructed by applying random transformations to the support pair. In this way, they enable training of the network without using annotations, i.e. the model is trained unsupervised. Finally, in the inference phase, they only need a few labeled image slices to perform segmentation on new classes.

However, a general problem with prototypical FSS is the loss of local information caused by average pooling of features during prototype extraction. This is particularly problematic for spatially heterogeneous classes like the background class in medical image segmentation problems, which can contain any semantic class other than the foreground class. Previous metric-learning based works have addressed this issue by computing additional prototypes per class to capture more diverse features. \cite{liu2020part} clustered the features within each class to obtain \textit{part-aware} prototypes and in the current state-of-the-art method, \cite{ouyang2020self} computed additional \textit{local} prototypes on a regular grid. 

We argue that it is insufficient to model the entire background volume with prototypes estimated from a few support slices and propose a conceptually different approach where we do \textit{not} increase the number of background prototypes but remove the need for these altogether. Inspired by the anomaly detection literature~\citep{chandola2009anomaly,ruff2021unifying}, we propose to only model the relatively homogeneous foreground class with a single prototype and introduce an anomaly score that measures the dissimilarity between this foreground prototype and all query pixels. Segmentation is then performed by thresholding the anomaly scores using a learned threshold that encourages compact foreground representations. For direct comparison of our novel anomaly detection-inspired few-shot medical image segmentation method to that of  \cite{ouyang2020self} and other representative works, our baseline setup follows their approach, working with 2D image slices. Within the existing 2D setup, we, as an added contribution, propose a new self-supervision task by extending the superpixel-based self-supervision scheme by~\cite{ouyang2020self} to 3D in order to utilize the volumetric nature of the data. As a natural extension, facilitated by the new self-supervision task, we further indicate potential benefits beyond this 2D setup by exploring a direct 3D treatment of the problem by employing a 3D convolutional neural network (CNN) as embedding network. 

By only explicitly modeling the foreground class, we argue that our proposed approach is more robust to background outside the support slices, compared to current state-of-the-art methods~\citep{ouyang2020self,roy2020squeeze}. To further illustrate this, we introduce a new evaluation protocol where we, based on labeled slices from the support image, segment the entire query image, thus being more exposed to background effects. Previous works, on the other hand, limit the evaluation of the query image only to the slices containing the class of interest. However, this approach requires additional weak labels in the form of information about the location of the class in the query image, which is unrealistic and cumbersome, especially in the medical setting. 

In summary, the main contributions of this work are three-fold. We propose: 
\begin{enumerate}[(1)]
    \item A simple but effective anomaly detection-inspired approach to FSS that outperforms prior state-of-the-art methods and removes the need to learn a large number of prototypes.
    \item A novel self-supervision task that exploits the 3D structural information in medical images within the 2D setup and indicate the potential of training 3D CNNs for direct volume segmentation.
    \item A new evaluation protocol for few-shot medical image segmentation that does not rely on weak-labels and therefore is more applicable in practical scenarios.
\end{enumerate}

\section{Related Work}
\subsection{Few-Shot Meta-learning}
As opposed to classical supervised learning that specializes a model to perform one specific task by optimizing over training samples, few-shot meta-learning optimizes over a set of training tasks, with the goal of obtaining a model that can quickly adapt to new, unseen tasks. There exist various approaches to few-shot learning, including \romannumeral 1) learning to fine-tune~\citep{finn2017model,ravi2016optimization}, \romannumeral 2) sequence based~\citep{mishra2017simple,santoro2016meta}, and \romannumeral 3) metric-learning based approaches~\citep{vinyals2016matching,snell2017prototypical,nguyensen}. Due to its simplicity and efficiency, the latter category has recently received a lot of attention, and the models relevant for this paper build on this principle. \cite{vinyals2016matching} combined deep feature learning with non-parametric methods in the Matching Network, by performing weighted nearest-neighbor classification in the embedding space. They proposed to train the model in episodes where a small labeled support set and an unlabeled query image are mapped to the query label, making the model able to adapt to unseen classes without the need for fine-tuning. Whereas the Matching Network only performed one-shot image classification, \cite{snell2017prototypical} later proposed the Prototypical Network, which extended the problem to include few-shot classification. Based on the idea that there exists an embedding space, in which samples cluster around their class prototype representation, they proposed a simpler model with a shared encoder between the support and query set, and a nearest-neighbor prototype matching in the embedding space.     

\subsection{Few-Shot Semantic Segmentation}
Few-shot semantic segmentation extends few-shot image classification~\citep{vinyals2016matching,snell2017prototypical,nguyensen} to pixel-level classifications~\citep{shaban2017one,rakelly2018conditional,zhang2020sg,wang2019panet}, and the goal is to, based on a few densely labeled samples from one (or more) new class(es), segment the class(es) in a new image. A recent line of work builds on the ideas from the Prototypical Network by \cite{snell2017prototypical}, and can be roughly split into two groups: models where predictions are based directly on the cosine similarity between query features and prototypes in the embedding space~\citep{wang2019panet,liu2020part,ouyang2020self}, and models that find the correlation between query features and prototypes by employing decoding networks to get the final prediction~\citep{dong2018few,zhang2019canet,liu2020crnet,li2021adaptive,zhang2021self,tian2020prior}.

\cite{dong2018few} first adopted the idea of metric-learning based prototypical networks to perform few-shot semantic segmentation. They proposed a two-branched model: a prototype learner, learning class-wise prototypes from the labeled support set, and a segmentation network where the prototypes were used to guide the segmentation of the query image. Most relevant for this work, \cite{wang2019panet} argued that parametric segmentation generalizes poorly, and proposed the Prototype Alignment Network (PANet), a simpler model where the knowledge extraction and segmentation process is separated. By exploiting prototypes extracted from the semantic classes of the support set, they reduced the segmentation of the query image to a non-parametric pixel-wise nearest-neighbor prototype matching, thereby creating a new branch of FSS models. Building on PANet, \citep{liu2020part} addressed the limitation of reducing semantic classes to a simple prototype and proposed the Part-aware Prototype Network (PPNet), where each semantic class is represented by multiple prototypes to capture more diverse features. \cite{liu2020part} further adopted a semantic branch for parametric segmentation during training to learn better representations. \cite{ouyang2020self} adapted ideas from PANet to perform FSS in the medical domain. They addressed the major restricting factor preventing medical FSS, e.g the dependency on a large a set of annotated training classes. This barrier was overcome by the introduction of a superpixel-based self-supervised learning scheme, enabling the training of FSS networks without the need for labeled data. \cite{ouyang2020self} further introduced the Adaptive Local Prototype pooling enpowered prototypical Network (ALPNet) where additional \textit{local} prototypes are computed on a regular grid to preserve local information and enhance segmentation performance. 

A different approach to medical FSS was suggested by \cite{roy2020squeeze}, and was the first FSS model for medical image segmentation. Their proposed SE-Net employs squeeze and excite blocks~\citep{hu2018squeeze} in a two-armed architecture consisting of one conditioner arm, processing the support set, and one segmenter arm, interacting with the conditioner arm to segment the query image. However, this model is trained supervised, requiring a set of labeled classes for training.

Based on our experience, training a decoder in a self-supervised setting, where the training task (superpixel segmentation) differs from the inference task (organ segmentation), is challenging and leads to performance degradation. In this paper, we thus, partially inspired by the state-of-the-art model \citep{ouyang2020self}, build further on the branch initiated by~\cite{wang2019panet} to perform FSS in the medical domain. We propose a novel FSS model that, unlike previous approaches in this branch~\citep{wang2019panet,liu2020part,ouyang2020self}, does \textit{not} explicitly model the complex background class, but relies solely on one foreground prototype.

\subsection{Self-Supervised Learning}
When large labeled datasets are not available, self-supervision can be used to learn representations by training the deep learning model on an auxiliary task that is defined such that the label is \textit{implicitly} available from the data. A good auxiliary task should require high-level image understanding to be solved, thereby encouraging the network to encode this type of information. Commonly used auxiliary tasks include image inpaining~\citep{larsson2016learning,pathak2016context,zhang2016colorful}, contrastive learning~\citep{chen2020simple,misra2020self}, rotation prediction~\citep{komodakis2018unsupervised}, solving jigsaw puzzles~\citep{noroozi2016unsupervised}, and relative patch location prediction~\citep{doersch2015unsupervised}. 

In the medical domain, self-supervised learning (SSL) has been used to improve performance on other (main) tasks by exploiting unlabeled data in a multi-task learning setting~\citep{chen2019self,li2021novel} and to pre-train models before transferring them to new (main) tasks~\citep{bai2019self,zhu2020rubik,dong2021self,lu2021volumetric}. In~\cite{ouyang2020self}, SSL was used to train a FSS model completely unsupervised using a novel superpixel-based auxiliary task, removing the need for labeled data during training. We build on this work by extending the proposed self-supervision scheme to 3D supervoxels. 

\begin{figure*}[t]
\begin{center}
   \includegraphics[width=0.95\linewidth]{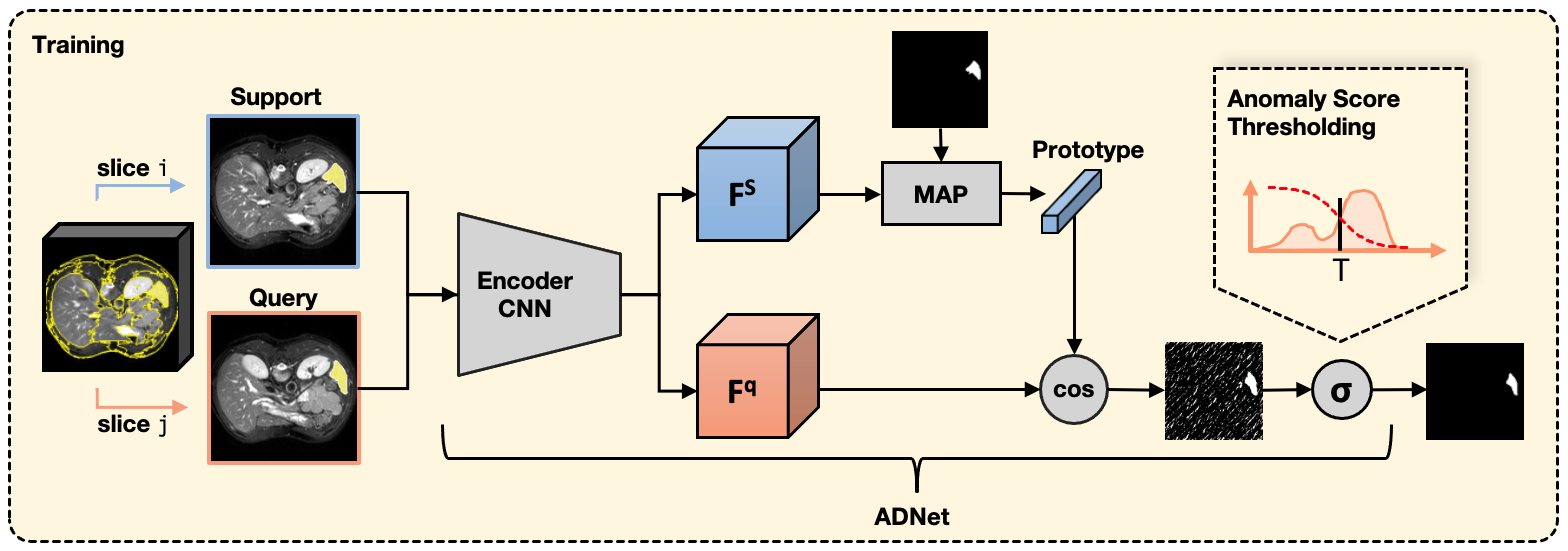}
\end{center}
  \caption{Illustration of the model during training. Support and query slices are obtained from the same image volume as two different 2D slices containing a randomly sampled supervoxel. A shared feature encoder encodes the query and the support images into deep feature maps. The support features are then resized to the mask size and masked average pooling is applied to compute the foreground prototype. For each query feature vector, an anomaly score is computed based on the cosine similarity to the prototype. Finally, the segmentation of the query image is performed by thresholding the anomaly scores using a learned anomaly threshold.}
\label{fig:training}
\end{figure*}

\begin{figure}[t]
\begin{center}
   \includegraphics[width=0.9\linewidth]{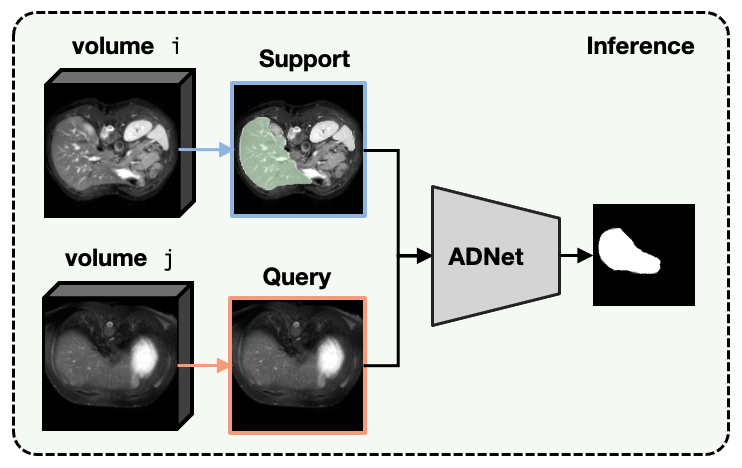}
\end{center}
  \caption{Illustration of the model during inference. Based on labeled slices from the support volume, the query volume is segmented slice by slice, one class at a time.}
\label{fig:inference}
\end{figure}

\subsection{Supervoxel Segmentation}
Supervoxels and superpixels are groupings of local voxels/pixels in an image that share similar characteristics. The boundaries of a supervoxel/superpixel therefore tend to follow the boundaries of the structures in the image, providing natural sub-regions. Supervoxel and superpixel segmentation has become a common tool in computer vision, also in the medical domain~\citep{huang2020segmentation,irving2016pieces}. For a detailed comparison of available superpixel segmentation algorithms, we refer the reader to~\citep{stutz2018superpixels}.

\section{Problem Definition}
Given a labeled dataset with classes $\mathcal{C}_{train}$ (here: $\mathcal{C}_{train} = \{supervoxel_1, supervoxel_2, ...\}$), FSS models aim to learn a quick adaption to new classes $\mathcal{C}_{test}$ (e.g. $\mathcal{C}_{test} = \{liver, kidney, spleen\}$) when exposed to only a few labeled samples. The training and testing are performed in an episodic manner \citep{vinyals2016matching} where, in each episode, $N$ classes are sampled from $\mathcal{C}$ to create a support set and a query set. The input to an episode is the support image(s) (with annotations) and a query image, and the output is the predicted query mask. In an $N$-way $k$-shot setting, the support set $\mathcal{S} = \left\{(\mathbf{x}_1, \mathbf{y}_1), ..., (\mathbf{x}_{N\times k}, \mathbf{y}_{N\times k}))\right\}$ consists of $k$ image slices $\mathbf{x} \in \mathbb{R}^{H\times W}$ (with annotations $\mathbf{y} \in \mathbb{R}^{H\times W}$ indicating the class of each pixel) from each of the $N$ classes, whereas the query set consists of one query image $\mathcal{Q} = \left\{(\mathbf{x}_1^*, \mathbf{y}_1^*)\right\}$ containing one or more of the $N$ classes. 

\section{Methods}
 In this work, we propose an anomaly detection-inspired network (ADNet) for prototypical FSS\footnote{By "anomaly" we refer to abnormalities compared to our defined normal class (foreground), and not necessarily something that occurs infrequently.}. We employ a shared feature extractor between the support and query images and perform metric learning-based segmentation in the embedding space. Unlike prior approaches that obtain prototypes for both foreground \textit{and} background classes~\citep{liu2020part,ouyang2020self,wang2019panet}, we only consider foreground prototypes to avoid the aforementioned problems related to explicitly modeling the large and heterogeneous background class. Based on \textit{one} foreground prototype, we compute anomaly scores for all query feature vectors. The segmentation of the query image is then based on these anomaly scores and a learned anomaly threshold. To train our model, we take inspiration from \cite{ouyang2020self} and propose a new supervoxel-based self-supervision pipeline. Fig.~\ref{fig:training} and Fig.~\ref{fig:inference} provide an overview of the model during training and inference, respectively.       

\subsection{Anomaly Detection-Inspired Few-Shot Segmentation} 
We denote the encoding network as $f_\theta$ and start by embedding the support and query images into deep features, $f_\theta(\mathbf{x}) = F^s$ and $f_\theta(\mathbf{x}^*) = F^q$, respectively. As opposed to previous works, we are only interested in explicitly modeling the foreground in each episode. We do this by employing the segmentation mask to perform masked average pooling (MAP), but only for the foreground class $c$. We resize the support feature map $F^s$ to the mask size ($H, W$) and compute one foreground prototype $p \in \mathbb{R}^d$, where $d$ is the dimension of the embedding space:

\begin{equation}
    p = \frac{\sum_{x,y} F^s(x, y) \odot {\bf y}^{fg}(x, y)}{\sum_{x,y} {\bf y}^{fg}(x, y)},
\end{equation}

\noindent where $\odot$ denotes the Hadamard product and ${\bf y}^{fg} = \mathbbm{1}( {\bf y} = c)$ is the binary foreground mask of class $c$\footnote{$\mathbbm{1}(\cdot)$ is the indicator function, returning 1 if the argument is true and 0 otherwise.}. 

To segment the query image based on this \textit{one} class-prototype, we design a threshold-based metric learning approach to the segmentation. We first obtain an anomaly score $S$ for each query feature vector $F^q(x, y)$ by calculating the (negative) cosine similarity to the foreground prototype $p$ of the episode:
\begin{equation}
    S(x, y) = - \alpha \frac{F^q(x, y) \cdot p}{\|F^q(x, y)\| \|p\|},
\label{eq:score}    
\end{equation}
where $\alpha = 20$ is a scaling factor introduced by \cite{oreshkin2018tadam}. In this way, query feature vectors that are identical to the prototype will get an anomaly score of $-\alpha$ (minimum), whereas query feature vectors that are pointing in the opposite direction, relative to the prototype, get an anomaly score of $\alpha$ (maximum). The predicted foreground mask is then found by thresholding these anomaly scores with a learned parameter $T$. To make the process differentiable, we perform soft thresholding by applying a shifted Sigmoid:
\begin{equation}
    {\bf\hat{y}}_{fg}^q(x, y) = 1-\sigma\left(S(x, y)-T\right),
    \label{eq:anomaly_score}
\end{equation}
where $\sigma(\cdot)$ denotes the Sigmoid function with a steepness parameter $\kappa = 0.5$. The impact of the steepness parameter is examined in Section~\ref{sec:steepness}. In this way, query feature vectors with an anomaly score below $T$ (similar to the prototype) get a foreground probability above $0.5$, whereas query feature vectors with an anomaly score above $T$ (dissimilar to the prototype) get a foreground probability below $0.5$. The predicted background mask is finally found as ${\bf\hat{y}}_{bg}^q = 1 - {\bf\hat{y}}_{fg}^q$.

The predicted foreground and background masks for the query image are then upsampled to the image size ($H,W$) and we compute the binary cross-entropy segmentation loss:
\begin{equation}
\begin{split}
    \mathcal{L}_{S} = - \frac{1}{HW} \sum_{x, y} & {\bf y}_{bg}^q(x, y) \log({\bf\hat{y}}_{bg}^q(x, y)) \\ 
    + &{\bf y}_{fg}^q(x, y) \log({\bf\hat{y}}_{fg}^q(x, y)).
\end{split}
\end{equation}
In order to encourage a compact embedding of the foreground classes, we construct an additional loss term $\mathcal{L}_{T} = T / \alpha $ that minimizes the learned threshold. The effect of this loss component is examined in Section~\ref{sec:ablation}.

Following common practice~\citep{liu2020part,ouyang2020self,wang2019panet}, we also add a prototype alignment regularization loss where the roles of support and query are reversed. The \textit{predicted} query mask is used to compute a prototype that segments the support image:
\begin{equation}
\begin{split}
    \mathcal{L}_{PAR} = - \frac{1}{HW} \sum_{x, y} & {\bf y}_{bg}^s(x, y) \log({\bf\hat{y}}_{bg}^s(x, y)) \\
    + & {\bf y}_{fg}^s(x, y) \log({\bf\hat{y}}_{fg}^s(x, y)).
\end{split}
\end{equation}
This gives us the overall loss function
\begin{equation}
    \mathcal{L} = \mathcal{L}_{S} + \mathcal{L}_{T} + \mathcal{L}_{PAR}.
\end{equation}

\subsection{Supervoxel-Based Self-Supervision}
The ADNet is parameterized by $\mathcal{P} = \{\theta, T\}$ and trained self-supervised (unsupervised) end-to-end in an episodic manner. For ease of comparison to previous approaches, our baseline setup follows a 2D approach, where volumes are segmented slice-by-slice. However, to better utilize the volumetric nature of the medical images, we propose a new self-supervision task that exploits 3D supervoxels during the model's training phase. As supervoxels are sub-volumes of the image, representing groups of similar voxels in local regions of the image volume, this allows us to sample 3D pseudo-segmentation masks for semantically uniform regions in the image.

In the training phase, each episode is constructed based on one unlabeled image volume and its supervoxel segmentation: First, one random supervoxel is sampled to represent the foreground class, resulting in a binary 3D segmentation mask. Then, we sample two 2D slices from the image containing this ”class”/supervoxel to serve as support and query images. By exploiting the relations across slices, we are able to increase the amount of information that can be extracted in the self-supervision task compared to prior approaches. Following \cite{ouyang2020self}, we additionally apply random transformations to one of the images (query or support) to encourage invariance to shape and intensity differences.

The supervoxels for all image volumes are computed offline using a 3D extension of the same unsupervised segmentation algorithm~\citep{felzenszwalb2004efficient} as in \citep{ouyang2020self}. This is an efficient graph-based image segmentation algorithm building on euclidean distances between neighboring pixels. In the 3D extension, this corresponds to the distances from each voxel to its $26$ nearest neighbours. In medical images, the resolution in $z$-direction (slice thickness) is typically different from the in-plane ($x,y$) resolution. To account for this anisotropic voxel resolution, we re-weight all distances along the $z$-direction ($xz-, yz-$ and $xyz-$direction) according to the spatial ratios. 

The supervoxel generation has one hyper-parameter $\rho$ that controls the minimum supervoxel size, where a larger $\rho$ corresponds to larger and fewer supervoxels. The effect of this parameter on the final segmentation result is examined in Section~\ref{sec:supervoxel_param}. 

\subsection{Implementation Details}
\label{sec:implementation}
The implementation is based on the PyTorch (v1.7.1) implementation of SSL-ALPNet~\citep{ouyang2020self}. The encoder network used in all the 2D experiments is a ResNet-101 pretrained on MS-COCO, where the classifier is replaced by a $1\times1$ convolutional layer to reduce the feature dimension from 2048 to 256. Following ALPNet, we optimize the loss using stochastic gradient descent with momentum 0.9, a learning rate of 1e-3 with a decay rate of 0.98 per 1k epochs, and a weight decay of 5e-4 over 50k iterations. To address the class imbalance, we follow previous work and weigh the foreground and background class in the cross-entropy loss (1.0 and 0.1, respectively). To further stabilize training, we set a minimum threshold of $200$ pixels on the supervoxel size in the slices sampled as support/query. Supervoxel generation is done offline (once per image volume) and is relatively computationally efficient\footnote{The compute time for generating all supervoxels for the MS-CMRSeg dataset is less than 3 minutes using a Quad-Core Intel Core i7 processor.}. Training takes $~1.8$h on a Nivida RTX 2080Ti GPU.

\section{Experiments}
\setlength{\textfloatsep}{13pt plus 0pt minus 3pt}
\subsection{Setup}
\subsubsection{Data} 
We assess the proposed method on representative publicly available datasets\footnote{Links to public datasets: \href{https://zmiclab.github.io/projects/mscmrseg19}{\color{blue}\rmfamily MS-CMRSeg} and \href{https://chaos.grand-challenge.org}{\color{blue}\rmfamily CHAOS}}: 
\begin{enumerate}[(1)]
    \item \textbf{MS-CMRSeg} (bSSFP fold), from the MICCAI 2019 Multi-sequence Cardiac MRI Segmentation Challenge, containing 35 3D cardiac MRI scans with on average 13 slices~\citep{zhuang2018multivariate,zhuang2016multivariate}.
    \item \textbf{CHAOS}, from the ISBI 2019 Combined Healthy Abdominal Organ Segmentation Challenge (task 5), containing 20 3D T2-SPIR MRI scans with on average 36 slices~\citep{CHAOS2021,CHAOSdata2019,kavur2019}.
\end{enumerate}

To compare our results to \cite{ouyang2020self}, we follow the same pre-processing scheme: 1) Cut the top 0.5\% intensities. 2) Re-sample image slices (short-axis slices for the cardiac images and axial slices for the abdominal images) to the same spatial resolution. 3) Crop slices to unify size ($256\times 256$ pixels). Further, to fit into the pretrained network, each slice is repeated three times along the channel dimension.    

\begin{figure}[t]
\begin{center}
   \includegraphics[width=0.95\linewidth]{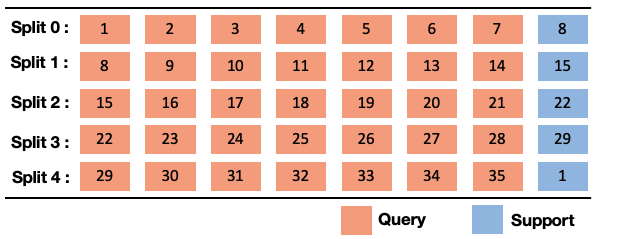}
\end{center}
  \caption{Setup for the five-fold cross-validation. This illustrates how the patient IDs are distributed among the splits and how the support/query volumes are selected for the cardiac MRI dataset. For each fold, a model is trained on all images \textit{not} present in that fold. During inference, the left-out fold is used exclusively, where the labeled support image is exploited to segment the query images slice by slice, class by class. The CHAOS dataset is split into five folds in a similar manner.}
\label{fig:splits}
\end{figure}

In all experiments, the models are trained self-supervised (unsupervised) and evaluated in a five-fold cross-validation manner, where, in each fold, the support images are sampled from \textit{one} of the patients and the remaining patients are treated as query (see Fig.~\ref{fig:splits}). Furthermore, to account for the stochasticity in the model and optimization, we repeat each fold three times. In the cardiac MRI scans we segment three classes: Left-ventricle blood pool (LV-BP), left-ventricle myocardium (LV-MYO) and right-ventricle (RV). In the abdominal MRI scans, we segment four classes: left kidney (L. kid.), right kidney (R. kid.), liver, and spleen. Following previous methods~\citep{ouyang2020self,roy2020squeeze}, each class is segmented separately in binary foreground/background segmentation problems\footnote{As the segmentation only relies on the computation of the cosine similarity to a class-specific prototype and a threshold which is shared among classes, the proposed method may be extended to account for multi-class scenarios. A detailed analysis of this is left for future work.}. Since the models are trained self-supervised, we do \textit{not} exclude image slices that contain the target classes.

\subsubsection{Evaluation metric} Following common practice \citep{ouyang2020self,roy2020squeeze} we employ the mean dice score to compare the model predictions to the ground truth segmentations. The dice score, $D$, between two segmentations $A$ and $B$ is given by 
\begin{equation}
    D(A,B) = 2\frac{|A\cap B|}{|A|+|B|} \cdot 100 \%,
    \label{eq:dice}
\end{equation}
meaning that a dice score of $100 \%$ corresponds to a perfect match between the segmentations.

\begin{figure}[t]
\begin{center}
   \includegraphics[width=.95\linewidth]{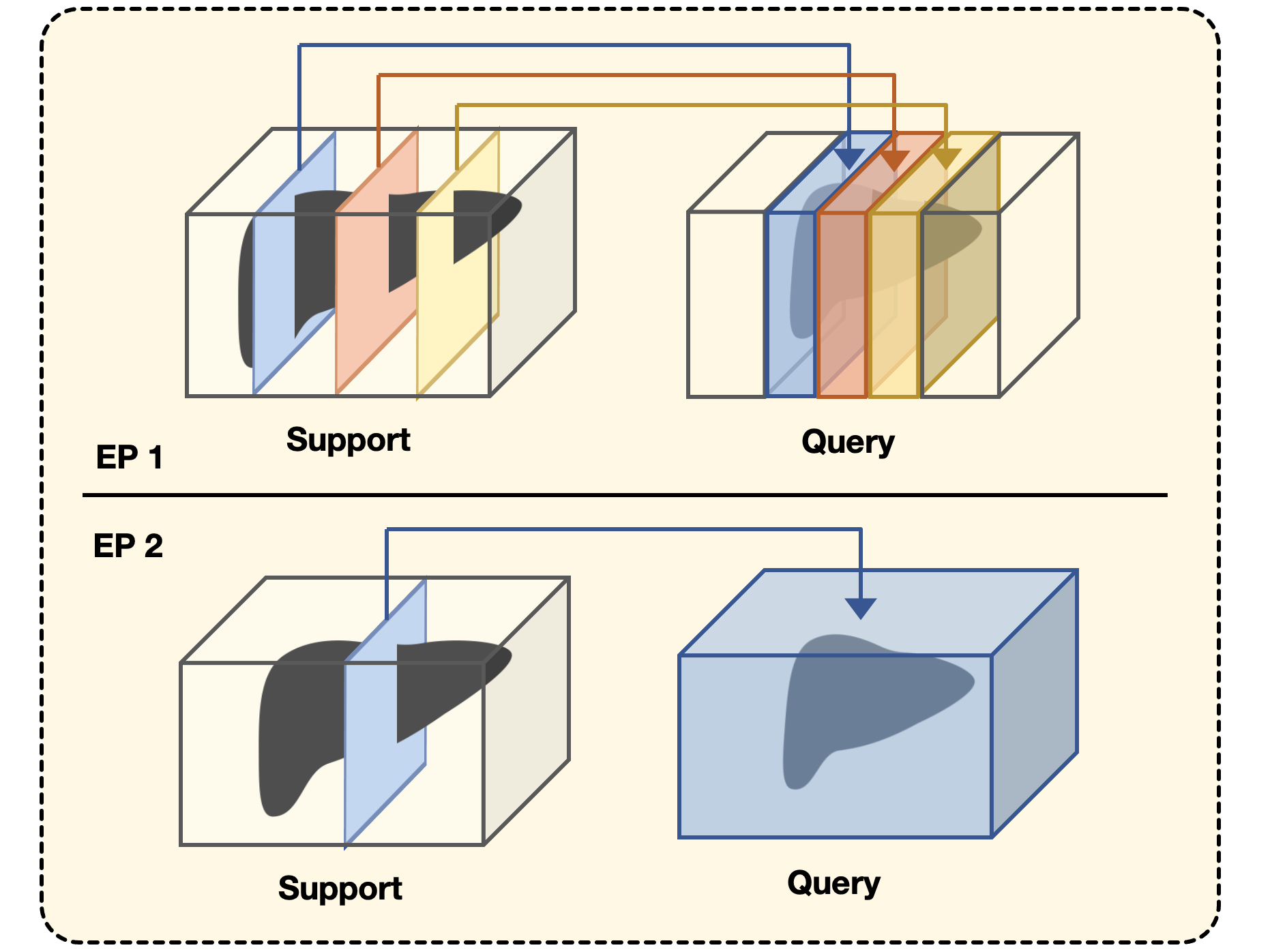}
\end{center}
  \caption{Illustration of EP1 (top) and EP2 (bottom). In EP1, the support and query volumes are divided into three succeeding  sub-chunks. The middle slice in each sub-chunk of the support volume is labeled and used to segment all the slices in the corresponding sub-chunk in the query volume. This means that the protocol requires weak labels indicating where the class of interest is located in the query volume. In EP2, the middle slice of the support volume is labeled and used to segment \textit{all} slices in the query volume, avoiding the need for additional weak labels.}
\label{fig:evaluation}
\end{figure}

\begin{table*}[t]
\setlength{\tabcolsep}{5pt} 
\renewcommand{\arraystretch}{1.2} 
\scriptsize
\caption{Mean dice score and standard deviation over three runs per split under EP1.}
\begin{center}
\begin{tabular}{lccc>{\color{orange}}clcccc>{\color{orange}}c}
\toprule
\multirow{2}{*}{Method} &
  \multicolumn{4}{c}{\textbf{Cardiac MRI}} &
   &
  \multicolumn{5}{c}{\textbf{Abdominal MRI}} \\ \cline{2-5} \cline{7-11} 
 &
  LV-BP &
  LV-MYO &
  RV &
  Mean &
   &
  L kid. &
  R kid. &
  Spleen &
  Liver &
  Mean \\ \hline
pSSL-PANet &
  $80.20\pm4.39$ &
  $45.67\pm2.58$ &
  $66.95\pm4.65$ &
  $64.27\pm14.23$ &
   &
  $63.09\pm9.31$ &
  $66.09\pm8.73$ &
  $63.93\pm8.65$ &
  $72.08\pm3.83$ &
  $66.30\pm3.51$ \\
pSSL-ALPNet &
  $\bm{87.54\pm1.63}$ &
  $60.19\pm4.55$ &
  $76.08\pm4.72$ &
  $74.60\pm11.21$ &
   &
  $\bm{81.00\pm4.01}$ &
  $\bm{84.66\pm2.40}$ &
  $72.32\pm7.69$ &
  $75.89\pm3.02$ &
  $78.46\pm4.72$ \\ \hdashline

vSSL-PPNet &
  $67.78\pm8.31$ &
  $42.61\pm6.16$ &
  $60.80\pm6.44$ &
  $57.06\pm10.61$ &
   &
  $62.13\pm7.85$ &
  $71.78\pm11.04$ &
  $66.57\pm9.04$ &
  $73.12\pm2.51$ &
  $68.40\pm4.37$ \\
vSSL-CANet &
  $78.99\pm4.72$ &
  $43.61\pm3.38$ &
  $61.10\pm3.60$ &
  $61.07\pm14.64$ &
   &
  $69.53\pm12.05$ &
  $77.15\pm10.71$ &
  $67.05\pm6.87$ &
  $72.88\pm3.27$ &
  $71.65\pm3.79$ \\
vSSL-ADNet &
  $87.53\pm2.03$ &
  $\bm{62.43\pm3.98}$ &
  $\bm{77.31\pm3.48}$ &
  $\bm{75.76\pm10.31}$ &
   &
  $75.28\pm14.80$ &
  $83.28\pm13.36$ &
  $\bm{75.92\pm8.90}$ &
  $\bm{80.81\pm2.36}$ &
  $\bm{78.82\pm3.35}$\\
\bottomrule
\end{tabular}
\end{center}
\label{tab:ep1}
\end{table*}

\begin{table*}[t]
\setlength{\tabcolsep}{5pt} 
\renewcommand{\arraystretch}{1.2} 
\scriptsize
\caption{Mean dice score and standard deviation over three runs per split under EP2. $^*$ indicates that the increase in mean dice score for the best performing model is statistically significant ($p < 0.05$).}
\begin{center}
\begin{tabular}{lccc>{\color{orange}}clcccc>{\color{orange}}c}
\toprule
\multirow{2}{*}{Method} &
  \multicolumn{4}{c}{\textbf{Cardiac MRI}} &
   &
  \multicolumn{5}{c}{\textbf{Abdominal MRI}} \\ \cline{2-5} \cline{7-11} 
 &
  LV-BP &
  LV-MYO &
  RV &
  Mean &
   &
  L kid. &
  R kid. &
  Spleen &
  Liver &
  Mean \\ \hline
pSSL-PANet &
  $68.28\pm5.67$ &
  $38.60\pm3.72$ &
  $55.22\pm5.18$ &
  $54.03\pm12.15$ &
   &
  $32.85\pm6.74$ &
  $30.18\pm4.85$ &
  $34.82\pm8.52$ &
  $53.89\pm3.15$ &
  $37.94\pm9.36$ \\
pSSL-ALPNet &
  $80.65\pm3.93$ &
  $53.31\pm6.31$ &
  $\bm{69.25\pm2.80}$ &
  $67.74\pm11.21$ &
   &
  $56.42\pm5.74$ &
  $50.37\pm7.77$ &
  $44.70\pm7.77$ &
  $56.73\pm3.07$ &
  $52.05\pm4.94$ \\ \hdashline

vSSL-PPNet &
  $56.69\pm8.35$ &
  $34.78\pm7.30$ &
  $47.60\pm6.07$ &
  $46.35\pm8.99$ &
   &
  $43.36\pm10.44$ &
  $56.94\pm14.39$ &
  $43.06\pm9.73$ &
  $56.32\pm7.57$ &
  $49.92\pm6.71$ \\
vSSL-CANet &
  $74.54\pm4.20$ &
  $35.08\pm4.30$ &
  $47.65\pm5.73$ &
  $52.42\pm16.46$ &
   &
  $50.18\pm13.02$ &
  $69.91\pm12.84$ &
  $48.84\pm8.61$ &
  $64.00\pm3.44$ &
  $58.23\pm8.98$ \\ 
vSSL-ADNet &
  $\bm{82.81\pm3.20}$ &
  $\bm{59.46\pm2.97}$ &
  $66.58\pm4.74$ &
  $\bm{69.62\pm9.77^*}$ &
   &
  $\bm{62.33\pm9.70}$ &
  $\bm{86.46\pm2.74}$ &
  $\bm{63.73\pm11.66}$ &
  $\bm{77.12\pm3.41}$ &
  $\bm{72.41\pm9.96^*}$ \\ \bottomrule
\end{tabular}
\end{center}
\label{tab:ep2}
\end{table*}

\subsubsection{Evaluation protocols}
During inference, the query volumes are segmented episode-wise, slice-by-slice, based on labeled support slices. For this reason, it is necessary to define an evaluation protocol that describes how to construct the episodes during inference, i.e. how to pair support and query images in episodes. In the experiments, we evaluate all models under two different evaluation protocols (EPs), illustrated in Fig.~\ref{fig:evaluation}. 

\paragraph{Evaluation protocol 1 (EP1)} Previous works~\citep{ouyang2020self,roy2020squeeze} follow an evaluation protocol that requires weak labels for all query images, i.e. there is a need to indicate (label) in which slices the foreground class is located. For a given class to be segmented, the chunk of slices in both the support and query volumes containing this class is divided into three succeeding sub-chunks. The middle slice in each sub-chunk of the support volume is used to segment all the slices in the corresponding sub-chunk in the query. In practice, this requires manual and time-consuming input from medical experts during the inference phase, where they have to scroll through each query image volume to mark the slices containing the class(es) of interest.  

\paragraph{Evaluation protocol 2 (EP2)} To avoid the need for weak query labels during inference, we introduce a new evaluation protocol that does not depend on the position of the target volume, and thus is more applicable in practical situations. Here, we simply sample $k=1$ slices from the support foreground volume and use this information to segment the entire query volume. To limit boundary effects, we choose the middle slice of the support foreground volume. 

\subsection{Comparison to state-of-the-art}
We compare our model to three modern FSS models: PANet~\citep{wang2019panet}, ALPNet~\cite{ouyang2020self}, and PPNet~\citep{liu2020part} with five (default) prototypes per class. Additionally, to compare our one-prototype anomaly approach to a one-prototype decoder approach, we adopt the dense comparison module proposed in~\citep{zhang2019canet} as a decoder on top of the backbone network and refer to this network as CANet\footnote{Code available: \href{https://github.com/kaixin96/PANet}{\color{blue}\rmfamily PANet}, \href{https://github.com/cheng-01037/Self-supervised-Fewshot-Medical-Image-Segmentation}{\color{blue}\rmfamily ALPNet},  \href{https://github.com/Xiangyi1996/PPNet-PyTorch}{\color{blue}\rmfamily PPNet}, and \href{https://github.com/icoz69/CaNet}{\color{blue}\rmfamily CANet}.}. 

The current state-of-the-art method for medical FSS, \cite{ouyang2020self}, showed that training PANet and ALPNet in a self-supervised manner improved the dice scores of the segmentation results considerably, compared to classical supervised FSS. Specifically, the dice scores on the MS-CMRSeg and CHAOS datasets increased by an average of $17.9$ and $26.1$ percentage points, respectively. Here, we are thus only focusing on SSL approaches. pSSL refers to the superpixel SSL approach presented in \cite{ouyang2020self}, whereas vSSL refers to our proposed supervoxel-based approach.

 \begin{figure*}[t]
\begin{center}
    \includegraphics[width=0.91\linewidth, trim=1.2cm 2cm 1.5cm 2cm, clip
    ]{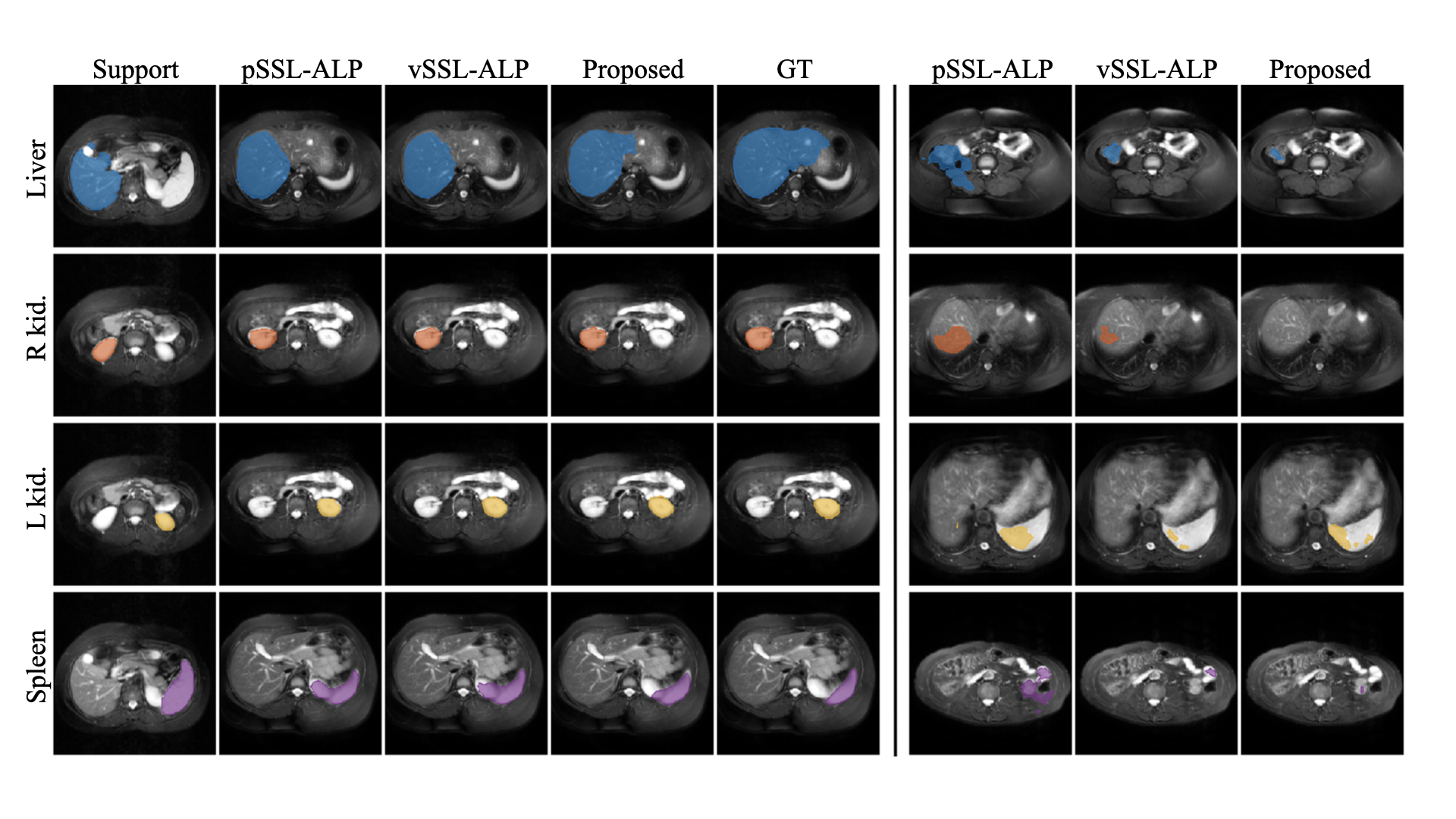}

\end{center}
  \caption{Qualitative comparisons for the abdominal MRI dataset. To the left of the solid line, we see (left to right) the support image, the segmentation results of a query slice containing the foreground class, and the ground truth segmentation of this query image. To the right, we see segmentation results for query slices \textit{not} containing the foreground class. Top to bottom: liver, right kidney, left kidney, and spleen. The proposed method is more robust to background outside the support slice, resulting in less over-segmentation.}
\label{fig:result_abd}
\end{figure*}

\begin{figure*}[!h]
\begin{center}
   \includegraphics[width=0.96\linewidth]{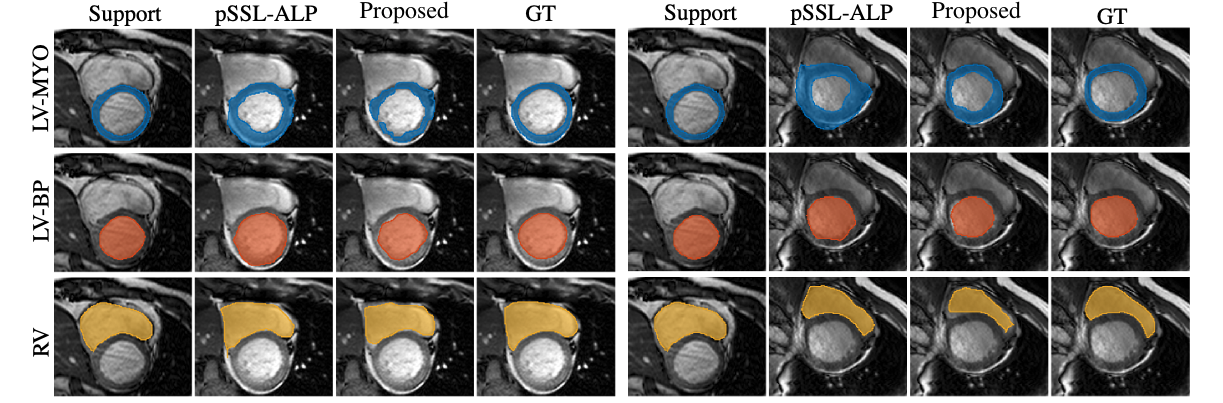}
\end{center}
  \caption{\textbf{Qualitative comparisons for two episodes with the same support volume from the cardiac MRI dataset}. Left to right: Support image, segmentation results of a query slice, and ground truth segmentation of this query image. The segmentation results are quite similar but the proposed method captures the left-ventricle myocardium and left ventricle blood pool better, with less over-segmentation.}
\label{fig:result_car}
\end{figure*}

\definecolor{LightCyan}{rgb}{0.88,1,1}
\begin{table}[t]
\setlength{\tabcolsep}{5pt} 
\renewcommand{\arraystretch}{1.2} 
\scriptsize
\caption{Summarized information about the models. $^*$The number of protototypes in ALPNet is adaptive and we report the average number over all classes during inference.}
\begin{center}
\begin{tabular}{lcccc}
\toprule
\multirow{2}{*}{Method} & \multirow{2}{*}{Backbone} & \multirow{2}{*}{Decoder} &\# Foreground & \# Background  \\
 & & & prototypes & prototypes \\
\hline

pSSL-PANet &  \multirow{5}{*}{ResNet-101} & \xmark & 1 & 1 \\

pSSL-ALPNet$^*$ &  & \xmark & 4 & 246  \\ 

vSSL-PPNet &  & \xmark & 5 & 5  \\

vSSL-CANet &  & \cmark & 1 & 0 \\ 

vSSL-ADNet &  & \xmark & 1 & 0 \\ 

\bottomrule
\end{tabular}
\end{center}
\label{tab:model_summary}
\end{table}

Table~\ref{tab:ep1} and Table~\ref{tab:ep2} present the results under EP1 and EP2, respectively, as mean and standard deviations over three runs (over all splits). Summarized details about the models can be found in Table~\ref{tab:model_summary}.

In Table~\ref{tab:ep1} we can see that our proposed model under EP1 performs similarly to the state-of-the-art on both datasets, while using significantly fewer prototypes compared to the closest competitors. We can also observe that the models that use just a few prototypes to model the background (PANet, PPNet) perform poorly and are among the three worst performing models for both datasets. Furthermore, by only modeling the foreground class and segmenting the query image using a decoding network, CANet results in the lowest (overall) dice score on the cardiac dataset. 
 
In a more realistic scenario, information about the location of the foreground volume in the query images is typically not available. We therefore evaluate the models under EP2 (Table~\ref{tab:ep2}) and we observe that our proposed approach outperforms the state-of-the-art. One-sided Wilcoxon signed rank tests~\citep{wilcoxon1992individual} on the mean dice scores across all runs indicate a significant difference between the segmentation results obtained from vSSL-ADNet and pSSL-ALPNet for both datasets under EP2 ($p < 0.05$). For the abdominal data, our model improves the segmentation results by more than 20 percentage points compared to pSSL-ALPNet. The main reason for this large improvement is that we now have to consider \textit{all} the query slices (not only the slices containing the organ to be segmented), meaning that the background class is much larger and much more diverse. This again complicates the task of modeling the background with prototypes, whereas our anomaly detection-inspired model without background prototypes is less affected. The somewhat lower performance and high standard deviation for left-kidney and spleen are related to the weak boundaries between these organs (see discussion in Section~\ref{sec:limit}). Furthermore, we obtain considerable, but smaller, improvements on the cardiac dataset under EP2. This is related to the lower number of slices and the less diverse background in these images, making the task of modeling the background with prototypes less complicated. Qualitative comparisons are provided in Fig.~\ref{fig:result_abd} and Fig.~\ref{fig:result_car}, where we can see that our approach is less prone to over-segmentation. 

\begin{figure}[t]
\begin{center}
   \includegraphics[width=\linewidth]{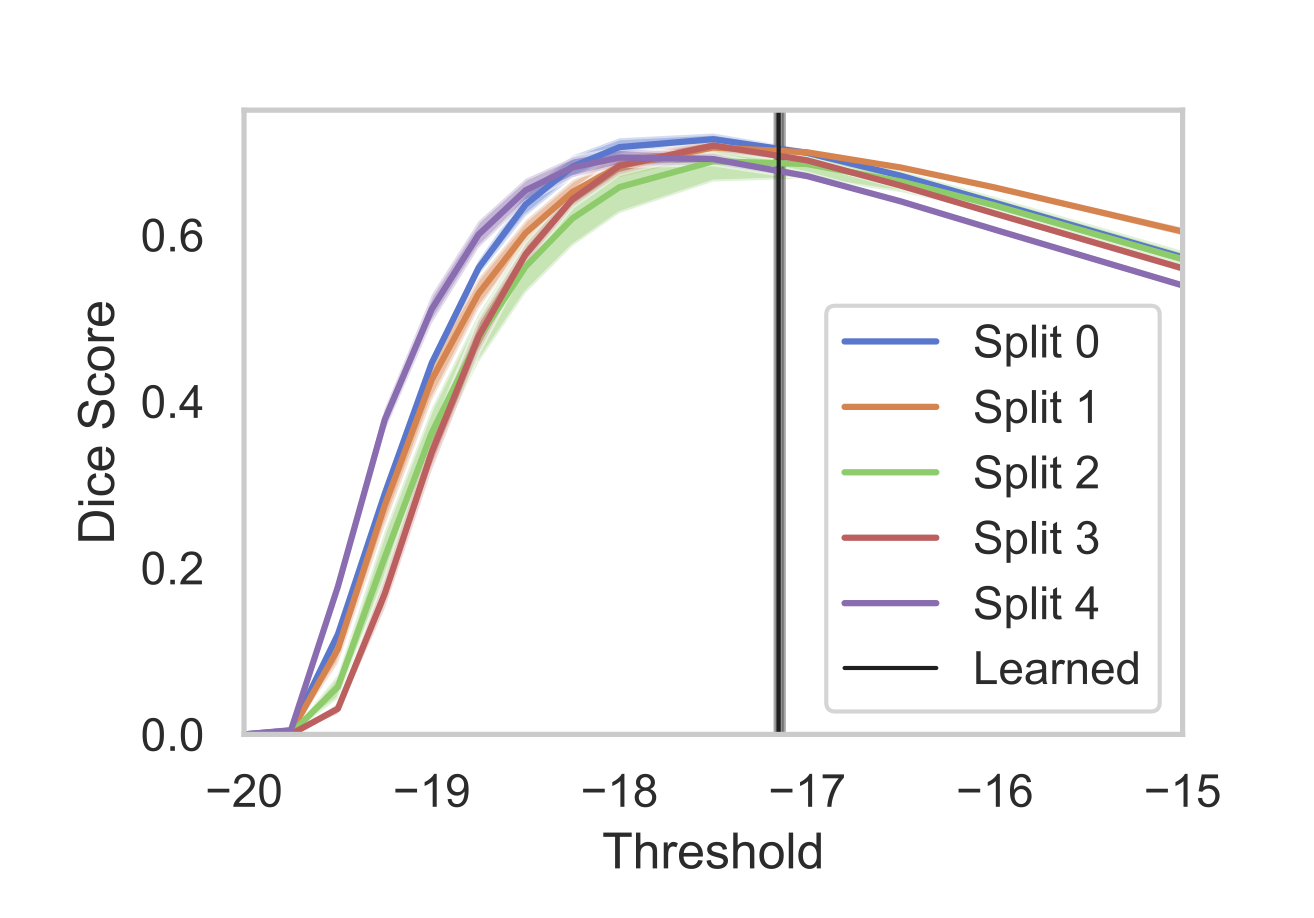}
\end{center}
  \caption{Analysis of the precision of the learned threshold. The plot shows the mean dice score (with standard deviation) obtained for a range of thresholds during inference on the MS-CMRSeg dataset. The learned threshold is indicated by the black vertical line.}
\label{fig:threshold_linesearch}
\end{figure}

\begin{table}[t]
\setlength{\tabcolsep}{5pt} 
\renewcommand{\arraystretch}{1.2} 
\scriptsize
\caption{Ablation study showing how the loss function components affect the results under EP1. $^*$ indicates that the increase in mean dice score for the best performing model is statistically significant ($p < 0.05$).}
\begin{center}
\begin{tabular}{cccccc>{\color{orange}}c}
\toprule
\multirow{2}{*}{$\mathcal{L}_{S}$} & \multirow{2}{*}{$\mathcal{L}_{T}$} & \multirow{2}{*}{$\mathcal{L}_{PAR}$} & \multicolumn{4}{c}{\textbf{Cardiac MRI}} \\
&  &  & LV-BP & LV-MYO & RV & Mean  \\
\hline
\cmark & \cmark & \cmark & $\bm{87.53\pm2.03}$ & $\bm{62.43\pm3.98}$ & $\bm{77.31\pm3.48}$ & $\bm{75.76\pm10.31^*}$\\
\cmark &  & \cmark & $87.41\pm2.08$ & $58.48\pm3.17$ & $74.95\pm3.33$ & $73.61\pm11.85$ \\
\cmark & \cmark & & $82.80\pm3.26$ & $57.70\pm3.05$ & $72.63\pm2.43$ & $71.05\pm10.31$  \\
\cmark &  &  & $83.62\pm2.51$ & $51.38\pm2.52$ & $68.36\pm3.02$ & $67.77\pm13.17$  \\

\bottomrule
\end{tabular}
\end{center}
\label{tab:ablation}
\end{table}

\begin{figure}[t]
\begin{center}
   \includegraphics[width=0.9\linewidth, trim=0 1cm 0 0, clip]{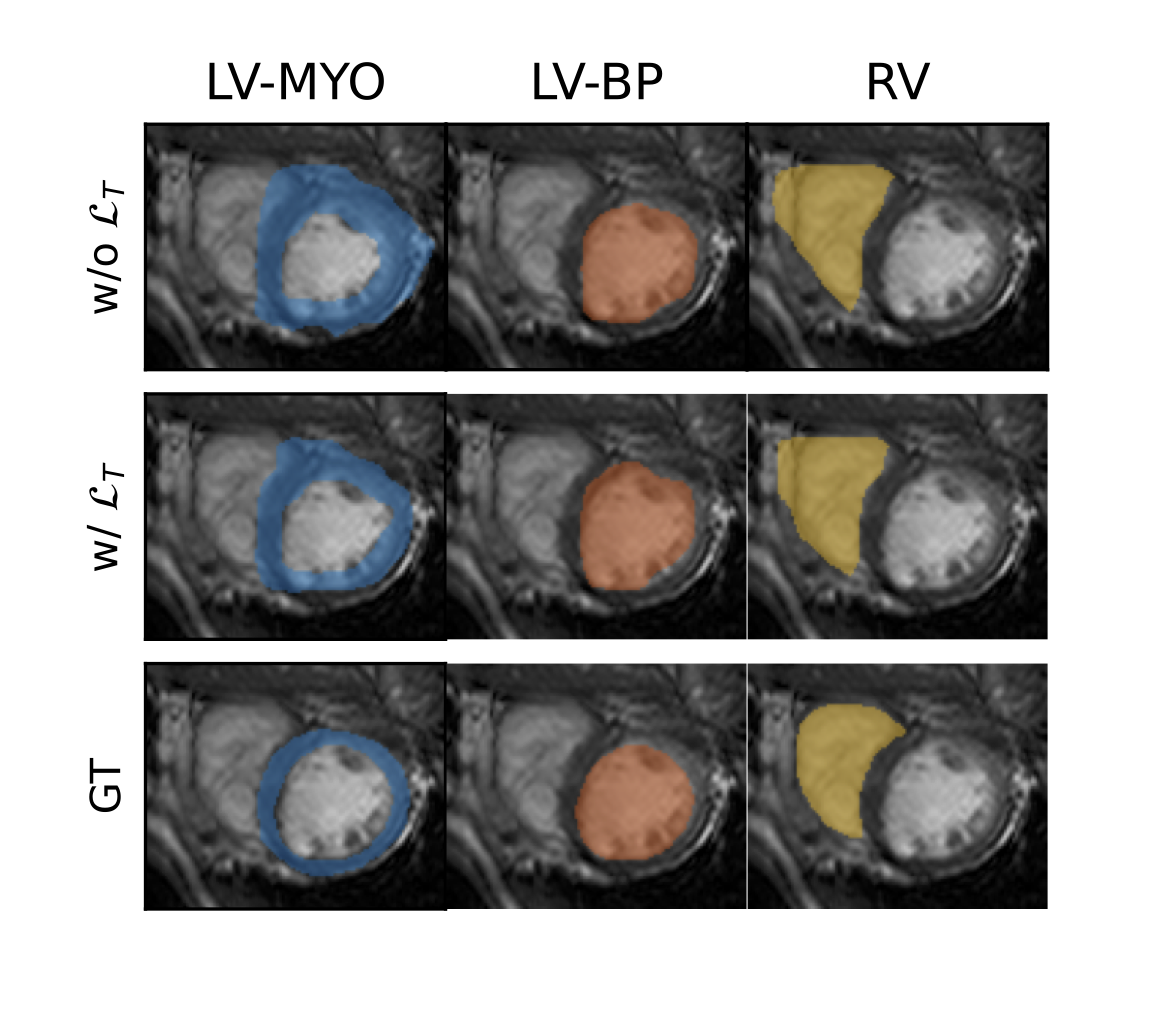}
\end{center}
  \caption{Qualitative (zoomed in) segmentation results for one slice in the MS-CMRSeg dataset obtained from a model trained with (middle) and without (top) $\mathcal{L}_{T}$ in the total loss. The lower row shows the ground truth, and it is evident that the threshold loss reduces the over-segmentation, especially for the left-ventricle myocardium.}
\label{fig:ablation}
\end{figure}

\subsection{Model analysis}
\subsubsection{Analysis of learned threshold}
To evaluate the learned threshold's precision on the unseen test data, we have conducted a line search where we, in the inference phase, evaluate the dice score obtained using a range of different thresholds between -20 and -15. The experiment was performed on three runs for each split and the mean dice score and standard deviation (shaded region) are reported in Fig.~\ref{fig:threshold_linesearch}. The learned threshold is averaged over all runs and and represented by the vertical black line\footnote{The small, gray shaded region indicates the range of learned threshold values.}. From the plot, we see that the threshold optimized for the training data is close to the ideal threshold for the test data, with little to gain in terms of increased dice score.

\begin{figure*}[t]
\begin{center}
   \includegraphics[width=0.98\linewidth]{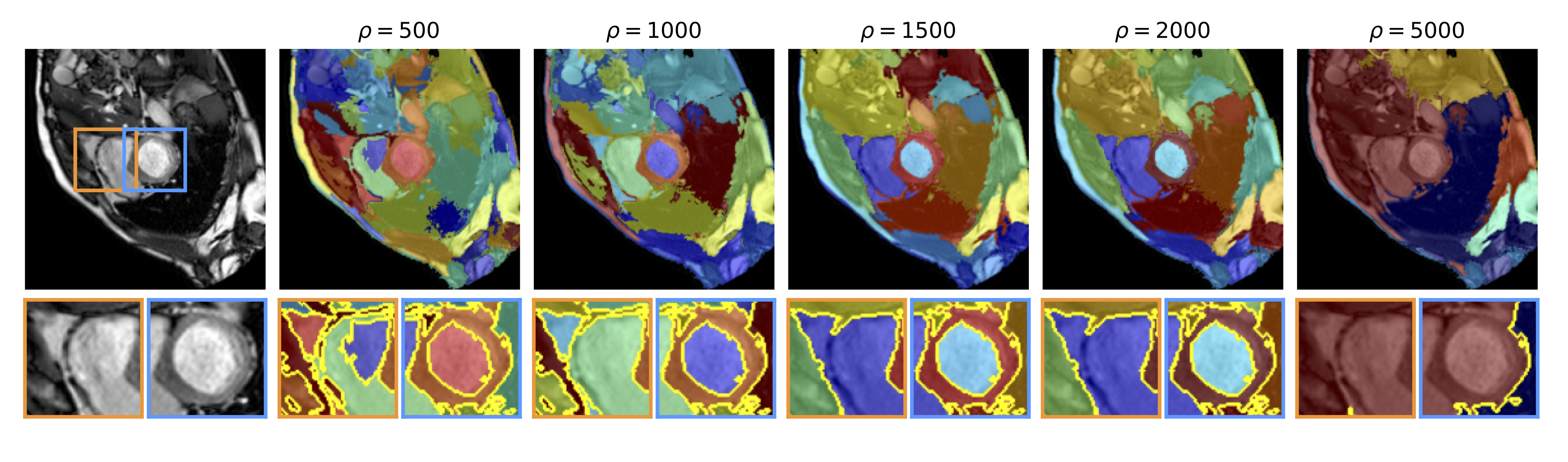}
\end{center}
  \caption{Examples of supervoxel segmentation results in one slice from the MS-CMRSeg dataset for different values of $\rho$. The parameter $\rho$ controls the minimum size of a supervoxel for it not to be joined with an adjacent supervoxel. A larger $\rho$ corresponds to larger and fewer supervoxels.}
\label{fig:supervoxels}
\end{figure*}

\subsubsection{Ablation study}
\label{sec:ablation}
To evaluate the effect of the three components of our loss function, we conduct an ablation study on the cardiac dataset. Table~\ref{tab:ablation} illustrates that $\mathcal{L}_{T}$ and $\mathcal{L}_{PAR}$ improve the dice score across all classes. Further, Fig.~\ref{fig:ablation} shows qualitatively the effect of $\mathcal{L}_{T}$ on the segmentation of one image slice from the MS-CMRSeg dataset. Here, it can be seen how the encouraging of a more compact foreground embedding via $\mathcal{L}_{T}$ reduces the over-segmentation, especially for the left-ventricle myocardium. 

\subsubsection{Sensitivity of supervoxel size}
\label{sec:supervoxel_param}
A sensitivity analysis of the parameter $\rho$, controlling the supervoxel size, is conducted on the MS-CMRSeg dataset and the results are presented in Table~\ref{tab:sensitivity}. As shown by these results, the final segmentation performance is relatively robust for a range of minimum size values from $\rho=1000$ to $\rho=2000$. However, if we allow the sizes to become too small ($\rho=500$) or too large ($\rho=5000$), we see that the performance is negatively affected. Examples of 2D slices from the 3D supervoxel segmentations for the different values of $\rho$ are shown in Fig.~\ref{fig:supervoxels}. 

According to the sensitivity study, a reasonable value is $\rho=1000$, and all the reported vSSL results are obtained with this value for the MS-CMRSeg dataset and $\rho = 5000$ for the CHAOS dataset, unless otherwise stated. The difference in value of $\rho$ reflects the differences in volume size.

\begin{table}[t]
\setlength{\tabcolsep}{6pt} 
\renewcommand{\arraystretch}{1.2} 
\scriptsize
\caption{\textbf{Supervoxel parameter sensitivity.} Analysis of the parameter controlling the minimum supervoxel size (n.o. voxels), on the cardiac MRI dataset under EP1.}
\begin{center}
\begin{tabular}{lccc>{\color{orange}}c}
\toprule
\multirow{2}{*}{$\rho$} & \multicolumn{4}{c}{\textbf{Cardiac MRI}} \\ 
& LV-BP & LV-MYO & RV & Mean \\
\hline
500 & $84.64\pm1.51$ & $43.48\pm6.25$ & $66.87\pm5.00$ & $65.00\pm16.86$ \\
1000 & $\bm{87.53\pm2.03}$ & $62.43\pm3.98$ & $\bm{77.31\pm3.48}$ & $\bm{75.76\pm10.31}$  \\
1500 & $86.91\pm2.47$ & $\bm{62.60\pm3.44}$ & $75.30\pm1.91$ & $74.94\pm9.93$ \\
2000 & $87.30\pm1.80$ & $61.21\pm3.33$ & $73.92\pm2.51$ & $74.14\pm10.65$ \\
5000 & $77.84 \pm 8.49$ & $49.30 \pm 7.93$ & $66.44 \pm 10.1$ & $64.53 \pm 14.72$\\
\bottomrule
\end{tabular}
\end{center}
\label{tab:sensitivity}
\end{table}

\subsubsection{Influence of steepness parameter}
\label{sec:steepness}
The steepness of the sigmoid function controls how soft the threshold operation performed is. If the steepness is high (harder thresholding), the class assignments of samples becomes harder, also close to the threshold. To examine the influence of the steepness parameter, $\kappa$, on the final segmentation results, we have conducted six experiments with different values of $\kappa$, from $\kappa = 0.1$ to $\kappa = 1.0$ on the MS-CMRSeg dataset\footnote{Note that this is equivalent to changing the scaling between $\alpha=0.2$ and $\alpha=20$ in Eq.~\eqref{eq:score}.}. The results presented in Table~\ref{tab:sigmoid_temp} indicate the model's robustness with respect to this parameter, and we can observe a gain of more than two percentage points in the dice score by decreasing the steepness from 1.0 to 0.5.

\begin{table}[t]
\setlength{\tabcolsep}{6pt} 
\renewcommand{\arraystretch}{1.2} 
\scriptsize
\caption{\textbf{Steepness parameter sensitivity.} Analysis of the parameter controlling the sigmoid steepness parameter, on the cardiac MRI dataset under EP1.}
\begin{center}
\begin{tabular}{llccc>{\color{orange}}c}
\toprule 
\multirow{2}{*}{$\kappa$} &
   &
  \multicolumn{4}{c}{\textbf{Cardiac MRI}} \\ 
 &
   &
  LV-BP &
  LV-MYO &
  RV &
  Mean \\ \hline
0.1 &
   &
  $80.69\pm4.04$ &
  $17.10\pm5.11$ &
  $52.69\pm10.35$ &
  $50.16\pm 26.02$ \\
0.3 &
   &
  $\bm{87.95\pm1.35}$ &
  $56.79\pm6.32$ &
  $\bm{78.44\pm2.48}$ &
  $74.39\pm13.04$ \\
0.5 &
   &
  $87.54\pm2.03$ &
  $\bm{62.44\pm3.98}$ &
  $77.32\pm3.48$ &
  $\bm{75.76\pm10.31}$ \\
0.7 &
   &
  $87.22\pm2.13$ &
  $62.21\pm2.71$ &
  $76.90\pm3.40$ &
  $75.45\pm10.26$ \\
0.9 &
   &
  $85.41\pm2.90$ &
  $60.67\pm3.57$ &
  $76.06\pm3.33$ &
  $74.05\pm10.20$ \\
1.0 &
   &
  $85.68\pm2.81$ &
  $59.40\pm4.04$ &
  $75.22\pm4.00$ &
  $73.43\pm10.80$ \\
\bottomrule 
\end{tabular}
\end{center}
\label{tab:sigmoid_temp}
\end{table}

\begin{table*}[t]
\setlength{\tabcolsep}{4pt} 
\renewcommand{\arraystretch}{1.2} 
\scriptsize
\caption{Mean dice score and standard deviation over three runs per split for ADNet and ALPNet with superpixel-based and supervoxel-based self-supevision. $^*$ indicates that the increase in mean dice score for the best performing model is statistically significant ($p < 0.05$).}
\begin{center}
\begin{tabular}{lccllll>{\color{orange}}llllll>{\color{orange}}l}
\toprule
\multirow{2}{*}{Model} &
  \multicolumn{1}{c}{\multirow{2}{*}{pSSL}} &
  \multicolumn{1}{c}{\multirow{2}{*}{vSSL}} &
   &
  \multicolumn{4}{c}{\textbf{Cardiac MRI}} &
   &
  \multicolumn{5}{c}{\textbf{Abdominal MRI}} \\ \cline{5-8} \cline{10-14} 
 &
  \multicolumn{1}{c}{} &
  \multicolumn{1}{c}{} &
   &
  \multicolumn{1}{c}{LV-BP} &
  \multicolumn{1}{c}{LV-MYO} &
  \multicolumn{1}{c}{RV} &
  \multicolumn{1}{>{\color{orange}}c}{Mean} &
   &
  \multicolumn{1}{c}{L kid.} &
  \multicolumn{1}{c}{R kid.} &
  \multicolumn{1}{c}{Spleen} &
  \multicolumn{1}{c}{Liver} &
  \multicolumn{1}{>{\color{orange}}c}{Mean} \\ \hline
\multirow{2}{*}{ALPNet} & \cmark &   &  & $\bm{80.65\pm3.93}$ & $53.31\pm6.31$ & $\bm{69.25\pm2.80}$ & $\bm{67.74\pm11.21}$ & & $56.42\pm5.74$ & $50.37\pm7.77$ & $44.70\pm7.77$ & $56.73\pm3.07$ & $52.05\pm4.94$  \\
                        &   & \cmark &  &  $79.44\pm2.79$ & $\bm{57.64\pm3.96}$ & $61.22\pm4.22$ & $66.10\pm9.59$ & & $\bm{68.19\pm12.30}$ & $\bm{82.45\pm6.27}$ & $\bm{55.39\pm10.09}$ & $\bm{66.38\pm5.20}$ & $\bm{68.10\pm9.62^*}$  \\\hdashline
\multirow{2}{*}{ADNet}  & \cmark &   &  &  $78.25\pm9.68$ & $54.59\pm6.42$ & $66.37\pm4.73$ & $66.40\pm12.07$ & & $49.65\pm7.59$ & $59.00\pm13.77$ & $52.47\pm9.56$ & $54.78\pm3.87$ & $53.97\pm 10.00$  \\
                        &   & \cmark &  &  $\bm{82.81\pm3.20}$ & $\bm{59.46\pm2.97}$ & $\bm{66.58\pm4.74}$ & $\bm{69.62\pm9.77^*}$ & & $\bm{62.33\pm9.70}$ & $\bm{86.46\pm2.74}$ & $\bm{63.73\pm11.66}$ & $\bm{77.12\pm3.41}$ & $\bm{72.41\pm9.96^*}$  \\
\bottomrule
\end{tabular}
\end{center}
\label{tab:ssl}
\end{table*}

\begin{table*}[t]
\setlength{\tabcolsep}{3pt} 
\renewcommand{\arraystretch}{1.5} 
\scriptsize
\caption{Mean dice score and standard deviation over three runs per split for vSSL-ADNet with 2D ResNet-101 as backbone and 3D ResNeXt-101 as backbone (under EP2). $^*$ indicates that the increase in mean dice score for the best performing model is statistically significant ($p < 0.05$).}
\begin{center}

\begin{tabular}{lcccccc>{\color{orange}}cccccc>{\color{orange}}c}
\toprule
\multirow{2}{*}{Backbone} & \multirow{2}{*}{Params} & \multirow{2}{*}{\begin{tabular}[c]{@{}l@{}}Labeled \\ slices, $k$\end{tabular}} & \multirow{2}{*}{} & \multicolumn{4}{c}{\textbf{Cardiac MRI}} &  & \multicolumn{5}{c}{\textbf{Abdominal MRI}} \\ \cline{5-8} \cline{10-14} 
 &  &  &  & LV-BP & LV-MYO & RV & Mean &  & L kid. & R kid. & Spleen & Liver & Mean \\ \hline
2D ResNet-101 & 42.50M & One &  & $82.81\pm3.20$ & $\bm{59.46\pm2.97}$ & $66.58\pm4.74$ & $\bm{69.62\pm9.77}$ &  & $62.33\pm9.70$ & $\bm{86.46\pm2.74}$ & $63.73\pm11.66$ & $77.12\pm3.41$ & $72.41\pm9.96$ \\
3D ResNeXt-101 & 47.52M & One &  & $81.28\pm2.51$ & $56.47\pm0.75$ & $66.22\pm4.24$ & $67.99\pm10.60$ &  & $77.95\pm16.57$ & $73.55\pm28.69$ & $75.04\pm8.55$ & $75.48\pm8.58$ & $75.50\pm17.71$ \\
3D ResNeXt-101 & 47.52M & All &  & $\bm{82.87\pm1.15}$ & $56.30\pm0.76$ & $\bm{67.93\pm4.04}$ & $69.03\pm11.15$ &  & $\bm{81.06\pm4.20}$ & $84.88\pm4.22$ & $\bm{75.18\pm8.40}$ & $\bm{77.17\pm8.60}$ & $\bm{79.58\pm7.67^*}$ \\
\bottomrule
\end{tabular}

\end{center}
\label{tab:3d}
\end{table*}

\subsubsection{vSSL vs pSSL}
\label{sec:vssl_pssl}
To disentangle and isolate the effect from the proposed extension of the self-supervision task, we have conducted additional experiments where we train our proposed model (ADNet), and the closest competing model (ALPNet) with the two different self-supervision tasks. From the results in Table~\ref{tab:ssl}, we see that the supervoxels overall yield better or comparable results for both models. For our proposed ADNet, there is a significant improvement ($p < 0.05$) in dice score from pSSL to vSSL for both datasets. Moreover, the improvements appear most prominent for the abdominal dataset, which is assumed to be related to the nature of the image volumes: In the abdominal dataset, the image volumes contain more slices and more potential information to utilize when the self-supervision task is extended to 3D, compared to the cardiac dataset.

A different implication of the proposed extension to supervoxel-based self-supervision is the enabling of training 3D CNNs for direct volume segmentation, as discussed in the next section.

\subsection{Extension to one-step volume segmentation}
\label{sec:3d}
Thus far, we have adopted a hybrid strategy to 3D segmentation, following~\cite{ouyang2020self}, where the 3D image volumes are segmented slice by slice, independently. However, a natural extension that is facilitated by the new self-supervision task is to adopt a 3D CNN as backbone to process the volumes in one step, thereby fully exploiting the potentially useful information along the third axis. Unfortunately, the high memory consumption and computational cost of 3D CNNs has limited their use to smaller images (in number of voxels), often obtained by down-sampling the original images~\citep{cciccek20163d} or by patch-based approaches~\citep{huo20193d}.

To investigate the potential of utilizing 3D convolutions to do one-step 3D segmentations within our proposed framework, we employ a 3D ResNeXt-101~\citep{hara2018can}, which is the 3D extension of ResNeXt~\citep{xie2017aggregated}, pretrained on the Kinetics-600 dataset~\citep{kay2017kinetics}, as our encoder network. The 3D ResNeXt-101 is a more resource efficient network, compared to the 3D ResNet-101, with approximately half as many trainable parameters in total. The number of parameters is comparable to the 2D ResNet-101 (see Table~\ref{tab:3d}).

To retain the same spatial resolution in the embedding space as for our 2D backbone, we modify the network by $i)$ removing the maxpooling in z-direction and $ii)$ changing the strides in conv 3, conv 4, and conv 5 to $(1, 2, 2)$, $(1, 1, 1)$, and $(1, 1, 1)$, respectively (see architecture details in Table~\ref{tab:architecture}). Similarly to the 2D ResNet-101, we replace the classifier with $1\times1\times1$ convolutions to reduce the feature dimension from 2048 to 256. Each voxel is repeated three times along the channel dimension in the input to fit into the pretrained network. The network is trained self-supervised end-to-end on 3D patches of size $(10, 215, 215)$, and the loss is optimized according to Section~\ref{sec:implementation}. During inference, we evaluate the performances under EP2 with two different levels of supervision: $i)$ Only labeling the middle slice of the target class in the support volume ($k = one$), as is done in the 2D experiments. $ii)$ Labeling all the support slices ($k = all$) and computing one prototype for the entire support volume, which is enabled by the volume-wise embedding.

Table~\ref{tab:3d} provides a summary of the performance of vSSL-ADNet with 3D ResNeXt-101 and 2D ResNet-101 backbones. Though it is difficult to directly compare 2D CNNs and 3D CNNs for many different reasons, such as difference in pre-training datasets and the number of weights modelling relations within slices and between slices, the results are meant to indicate the potential of using 3D convolutions in our framework to perform one-step 3D segmentation.

\begin{table}[t]
\setlength{\tabcolsep}{5pt} 
\renewcommand{\arraystretch}{1.4} 
\scriptsize
\caption{Modified 3D ResNeXt-101 architecture with cardinality $C=32$ used as backbone in the 3D experiments.}
\begin{center}
\begin{tabular}{p{0.15\linewidth}  p{0.14\linewidth} p{0.5\linewidth}}
\toprule
\textbf{Layer name} & \textbf{Output size} & \textbf{Architecture}  \\
\hline
conv 1 & ($1,\tfrac{1}{2}, \tfrac{1}{2}$) & $7\times 7\times 7$, 64, stride $1,2,2$ \\
\hline
\multirow{4}{*}{conv 2} & \multirow{4}{*}{($1,\tfrac{1}{4}, \tfrac{1}{4}$)} & $1\times3\times3$ max pool, stride $1,2,2$ \\
& & $\begin{bmatrix*}[l]
1\times1\times1, 128, $ stride $ 1,1,1 \\
3\times3\times3, 128, $ stride $ 1,1,1, C=32  \\
1\times1\times1, 256, $ stride $ 1,1,1
\end{bmatrix*} \times 3$\\
\hline
conv 3 & ($1,\tfrac{1}{8}, \tfrac{1}{8}$) & $\begin{bmatrix*}[l]
1\times1\times1, 256, $ stride $ 1,1,1 \\
3\times3\times3, 128, $ stride $ 1,2,2, C=32  \\
1\times1\times1, 512, $ stride $ 1,1,1
\end{bmatrix*} \times 4$\\
\hline
conv 4 & ($1,\tfrac{1}{8}, \tfrac{1}{8}$) & $\begin{bmatrix*}[l]
1\times1\times1, 512, $ stride $ 1,1,1 \\
3\times3\times3, 512, $ stride $ 1,1,1, C=32 \\
1\times1\times1, 1024, $ stride $ 1,1,1
\end{bmatrix*} \times 23$\\
\hline
conv 5 & ($1,\tfrac{1}{8}, \tfrac{1}{8}$) & $\begin{bmatrix*}[l]
1\times1\times1, 1024, $ stride $ 1,1,1 \\
3\times3\times3, 1024, $ stride $ 1,1,1, C=32 \\
1\times1\times1, 2048, $ stride $ 1,1,1
\end{bmatrix*} \times 3$\\
\hline
conv 6 & ($1,\tfrac{1}{8}, \tfrac{1}{8}$) & $1\times 1\times 1$, 256, stride $1,1,1$ \\
\bottomrule
\end{tabular}
\end{center}
\label{tab:architecture}
\end{table}

From the results on the cardiac dataset, we see that the differences between 2D and 3D are relatively small, which agrees with observations in previous work~\citep{vesal2019automated}. In the abdominal dataset, on the other hand, there appears to be a greater potential for utilizing the 3D structure via 3D convolutions. This mirrors our results from Section~\ref{sec:vssl_pssl}, where we found that the abdominal dataset benefited more from extending the self-supervision task from superpixels to supervoxels. 

\begin{figure}[t]
\begin{center}
   \includegraphics[width=0.9\linewidth]{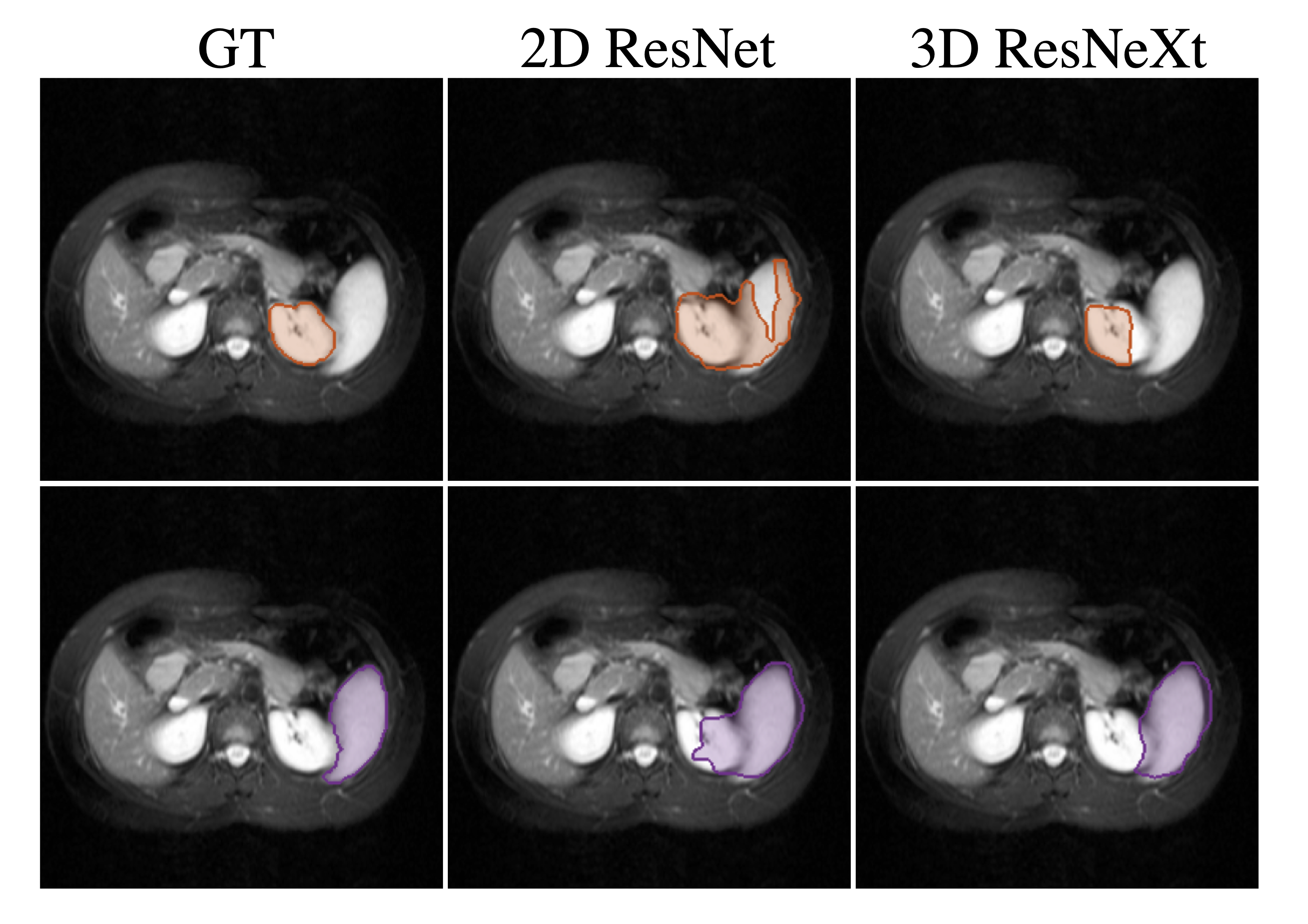}
\end{center}
  \caption{Comparison of the segmentation results for the left kidney (orange, top) and spleen (purple, bottom) classes for vSSL-ADNet with 2D ResNet-101 and 3D ResNeXt-101 as backbone. The 3D CNN leads to a better separation between the classes.}
\label{fig:lkrk}
\end{figure}

The largest performance difference between the backbones can be observed for the left kidney and spleen classes. While the 2D CNN results in a segmentation where these classes are confused, the 3D CNN leads to a better separation between the classes, as illustrated in Fig.~\ref{fig:lkrk}. We further observe a drop in performance on the right kidney class for the 3D CNN with $k=1$, which demonstrates the importance of having good support features to achieve robust results with the 3D backbone.

\section{Limitations and Outlook}
\label{sec:limit}
The key observation leading to our anomaly-detection inspired few-shot medical image segmentation is that the foreground class typically is relatively homogeneous. By only modeling the foreground class with a single prototype, we avoid having to model the large and highly inhomogeneous background, which we believe is the main challenge in prototypical few-shot medical image segmentation. However, if our assumption of a relatively homogeneous foreground class is not met, and the foreground consists of multiple distinct regions with strong edges, e.g. combining left-ventricle blood pool and left-ventricle myocardium into one foreground class (left-ventricle), modeling the foreground with one prototype might not be sufficient. This is related to the nature of the supervoxels, which tend to follow the boundaries of the structures in the image; Left-ventricle blood pool and left-ventricle myocardium will typically belong to different supervoxels during training and the network therefore learns to separate their feature representations into different clusters. To be able to capture this combined foreground class during inference, one option could be to take inspiration from PPNet~\citep{liu2020part} and cluster the features into multiple foreground prototypes and then merge the results.

Both the superpixel-based and the supervoxel-based self-supervision tasks are inevitably vulnerable to merging different classes during training \textit{if} the boundaries between them are weak: If the boundaries are weak, the classes will end up in the same superpixel/voxel and the network learns to embed the classes into the same cluster, which makes them difficult to separate during inference. Moreover, in the supervoxel case, it is enough for \textit{one} slice to contain a weak boundary between the classes before they leak into the same supervoxel. This is something that happens between the left-kidney and the spleen in the abdominal dataset, and leads to confusion between these two classes during inference, thereby resulting in lower dice scores and high standard deviations. Taking into account this weak/noisy nature of the supervoxel pseudo-labels is a promising direction for future research.

\section{Conclusion}

In this work, we proposed a novel and end-to-end trainable anomaly detection-inspired FSS network for medical image segmentation. By approaching the segmentation task as an anomaly detection problem, our model eliminates the need to explicitly model the large and heterogeneous background class. Moreover, to train the model in an unsupervised manner, we introduced a new self-supervision task that captures the 3D nature of the data by utilizing supervoxels. We assessed our proposed model on representative datasets for cardiac segmentation and abdominal organ segmentation, and showed that it improves segmentation performance and robustness, especially in the realistic scenario where no weak labels for the query images are assumed. Furthermore, we demonstrated how the proposed model, together with the new self-supervision task, has the potential to perform one-step 3D segmentation of the entire image volumes. We believe that fully exploiting the 3D nature of the medical images in this manner for few-shot segmentation represents an interesting line of research for future work.

\section*{Acknowledgements}
This work was supported by The Research Council of Norway (RCN), through its Centre for Research-based Innovation funding scheme [grant number 309439] and Consortium Partners; RCN FRIPRO [grant number 315029]; RCN IKTPLUSS [grant number 303514]; and the UiT Thematic Initiative. 

\bibliographystyle{model2-names.bst}\biboptions{authoryear}
\bibliography{refs}


\end{document}